\documentclass[10pt, paper=a4, UKenglish]{article}
\usepackage{graphicx}
\def\Title#1{\begin{center} {\Large #1 } \end{center}}
\def\Author#1{\begin{center}{ \sc #1} \end{center}}
\def\Address#1{\begin{center}{ \it #1} \end{center}}

\newcommand\pubblock{\rightline{\begin{tabular}{l} Proceedings of the CTD/WIT 2019\\ \pubnumber\\
         \pubdate  \end{tabular}}}

\newenvironment{Abstract}{\begin{quotation} \begin{center} 
             \large ABSTRACT \end{center}\bigskip 
      \begin{center}\begin{large}}{\end{large}\end{center} \end{quotation}}

\newenvironment{Presented}{\begin{quotation} \begin{center} 
             PRESENTED AT\end{center}\bigskip 
      \begin{center}\begin{large}}{\end{large}\end{center} \end{quotation}}

\def\beq{\begin{equation}}
\def\eeq#1{\label{#1}\end{equation}}
\def\eeqn{\end{equation}}

\def\beqa{\begin{eqnarray}}
\def\eeqa#1{\label{#1}\end{eqnarray}}
\def\eeqan{\end{eqnarray}}

\let\bar=\overbar

\def\Dslash{\not{\hbox{\kern-4pt $D$}}}
\def\dslash{\not{\hbox{\kern-2pt $\del$}}}

\def\msb{{\bar{\ssstyle M \kern -1pt S}}}

\usepackage{color}
\usepackage{lineno}
\usepackage{subfig}
\usepackage{hyperref}

\newcommand\pubnumber{PROC-CTD19-030}

\newcommand\pubdate{\today}

\def\affiliation{
On behalf of the CMS Muon Group, \\
Centro de Investigaciones Energ\'eticas, Medioambientales y Tecnol\'ogicas, Spain}

\newcommand{\conference}{Connecting the Dots and Workshop on Intelligent Trackers (CTD/WIT 2019)\\
Instituto de F\'isica Corpuscular (IFIC), Valencia, Spain\\ 
April 2-5, 2019}

\usepackage{fancyhdr}
\definecolor{mygrey}{RGB}{105,105,105}
\begin{document}


\large
\begin{titlepage}
\pubblock

\vfill
\Title{A firmware oriented trigger algorithm for CMS Drift Tubes in HL-LHC}
\vfill

\Author{Silvia Goy L\'opez}
\Address{\affiliation}
\vfill

\begin{Abstract}
A full replacement of the existing muon trigger system in the CMS (Compact Muon Solenoid) detector is planned for operating at the maximum instantaneous luminosities of about $5-7.5\times10^{34}$~cm$^{-2}$~s$^{-1}$ expected in HL-LHC (High Luminosity Large Hadron Collider). In this plan, the new on-detector electronics that is being designed for the DT (Drift Tubes) detector will read out all the chamber information at its maximum time resolution. A new trigger system, based on the highest performing FPGAs, is being designed and will be capable of providing precise muon reconstruction and bunch crossing identification. An algorithm easily portable to an FPGA architecture has been designed to implement the new trigger primitive information from the DT detector. This algorithm reconstructs muon segments from single-wire DT hits, which contain a time uncertainty of 400 ns due to the drift time in the cell. This algorithm provides the maximum resolution achievable by the DT chambers and brings the hardware system closer to the offline performance capabilities. The results of the simulation and of the first implementations in the new electronics test bench will be shown. 
\end{Abstract}

\vfill

\begin{Presented}
\conference
\end{Presented}
\vfill
\end{titlepage}
\def\thefootnote{\fnsymbol{footnote}}
\setcounter{footnote}{0}
%

\normalsize 


\section{Introduction}
\label{intro}
The Compact Muon Solenoid (CMS) is one of two general purpose detectors at the LHC. The central feature of the CMS apparatus is a superconducting solenoid of 6 m internal diameter, providing an axial magnetic field of 3.8 T. 
The CMS experiment has been designed with a two-level trigger system: the Level-1 Trigger (L1), implemented in custom- designed electronics; and the High Level Trigger (HLT), a streamlined version of the CMS offline reconstruction software running on a computer farm. 
Muon reconstruction in CMS is performed with the all-silicon tracker at the heart of the detector, and with up to four stations of gas-ionization muon detectors installed outside the solenoid and sandwiched between steel layers serving both as hadron absorbers and as a return yoke for the magnetic field. Drift Tube (DT) chambers and Cathode Strip Chambers (CSC) are used within $|\eta|<1.2$ and $0.9 <|\eta|< 2.4$ respectively, complemented by a system of Resistive Plate Chambers (RPC) covering the range of $|\eta|<1.6$.
A more detailed description of the CMS detector, together with a definition of the coordinate system used and the relevant kinematic variables, can be found in Ref\cite{CMSdes}.
 
The basic element of the CMS Drift Tube (DT) detector is the drift cell. The transverse size is $42\times13$ mm$^{2}$ with a 50~$\mu$m diameter gold-plated stainless steel anode wire at the centre. The gas is a mixture of Ar/CO, which provides a saturated drift velocity of about 54~$\mu$m/ns. The drift time of electrons produced by ionization is measured by the electronics, and the maximum drift time is $\sim$ 390~ns. Hits are then reconstructed, with a left-right ambiguity in their position at the layer level. Four staggered layers of parallel cells form a superlayer (SL), which allows for solving single-hit ambiguities and provides the measurement of two-dimensional segments. A chamber is composed by two superlayers measuring the r-phi coordinates (SL1, SL3), with the wires parallel to the beam line and separated by $>$20~cm, and an orthogonal superlayer measuring the r-z coordinates (SL2)

For the upcoming high-luminosity phase of the LHC, the current DT electronics will be replaced with a more robust and flexible solution in order to minimize possible effects of radiation damage and improve the intrinsic limitations of the current design.
The signal digitation happens inside the chamber where Time-to-Digital Converters (TDCs) with less than 1~ns resolution will be hosted in radiation-tolerant FPGAs. All digitised signals will be asynchronously streamed via optical links to the backend, outside of the experimental cavern. In the current system trigger primitives, that provide the relevant information from the DT to the CMS L1 system, are generated in on-detector custom electronics with limited time and spatial resolution~\cite{DTLT}. In the upgraded DT system all the high level functionalities, in particular, the trigger primitive generation, happens off-detector in powerful FPGAs.  This results in a substantial improvement of the physics performance at trigger level of less than 1~ns time resolution and no major regional constrains, i.e., comparable to current offline performance.
We present here a possible algorithm, called "Analytical Method", to perform trigger primitive generation for DT upgrade. This is being implemented both in software and firmware. Initial performance studies are shown here using the current readout chain to mimic the input to the Phase 2 electronics, due to their similar characteristics.

\section{Description of the Analytical Method}
\label{AMEmu}
The input information to the Analytical Method for trigger primitive generation is the wire position of the hit cells and the corresponding hit times from the start of the LHC orbit. From this, and assuming a given laterality, the hit positions can be reconstructed. For a given hypothesis of muon trajectory within a superlayer, which is a straight line, the collision time is determined from three cells, because the dependence on track slope is factored out. Thus, the bunch crossing (BX) of the corresponding proton-proton interaction that produced the muon is determined. In practice a selection is made of patterns of 4 tubes and their subpatterns of 3 tubes over 10 cells at a time, containing all physical trajectories in the given superlayer. For cases with 4 hits (one per layer), time is computed from each triplet and then combined in an arithmetic mean. Once the collision time is known, the track parameters are computed using exact formulas from a least squares method (chi2-minimisation). For groups of 3 hits all hit laterality assumptions providing physical solutions are considered as candidates. For groups of 4 hits a unique final candidate is selected, the one with minimum chi2. For muons with fits of 4 hits or 3 hits both in superlayer 1 (SL1) and and superlayer 3 (SL3), the information from both fits can be correlated if the corresponding segment times are within a window of +/- 25 ns. If a match is found, the candidate trigger primitive parameters are redefined as follows: the new time is the mean of the per superlayer fits times, the new position is the mean of the superlayer fits positions, and the new slope is computed from the difference in fit positions in SL3 and SL1 divided by distance between the two r-phi superlayers. If no match found, all per-superlayer candidates are kept. This algorithm has been implemented in CMS software (CMSSW) as an emulator for the firmware implementation in FPGA.

\section{Intrinsic and emulator performances}
\label{IntEmuPerf}
To assess the potential and the performance achieved using this algorithm, we have compared the reconstructed time and slope of the resulting trigger primitives with respect to the offline reconstructed segments for two possible workflows.
As first step, we have evaluated the performance for obtaining the track parameters with a least square fitting method (chi2 minimisation) for all variables, including the time for fits with 4 hits, in clean conditions, i.e. taking as input the calibrated times of only those hits associated to offline segments. This allows an evaluation of the intrinsic performance of a 'exact solution' for obtaining all the track parameters. In the following we refer to this as 'intrinsic performance', and results are shown in section \ref{IntPer}.
In the second workflow, the trigger primitive is reconstructed with the Analytical Method as implemented in the Emulator, including the full grouping and pattern recognition previously described. In case more than one trigger primitive is generated, the one closest to the offline segment in x coordinate is selected for performance studies . We refer to this as 'emulator performance', and the results are shown in section \ref{EmuPer}.

\section{Intrinsic Performance Studies}
\label{IntPer}
\begin{figure}[!htb]
  \centering
  \subfloat[]{\includegraphics[width=0.4\linewidth]{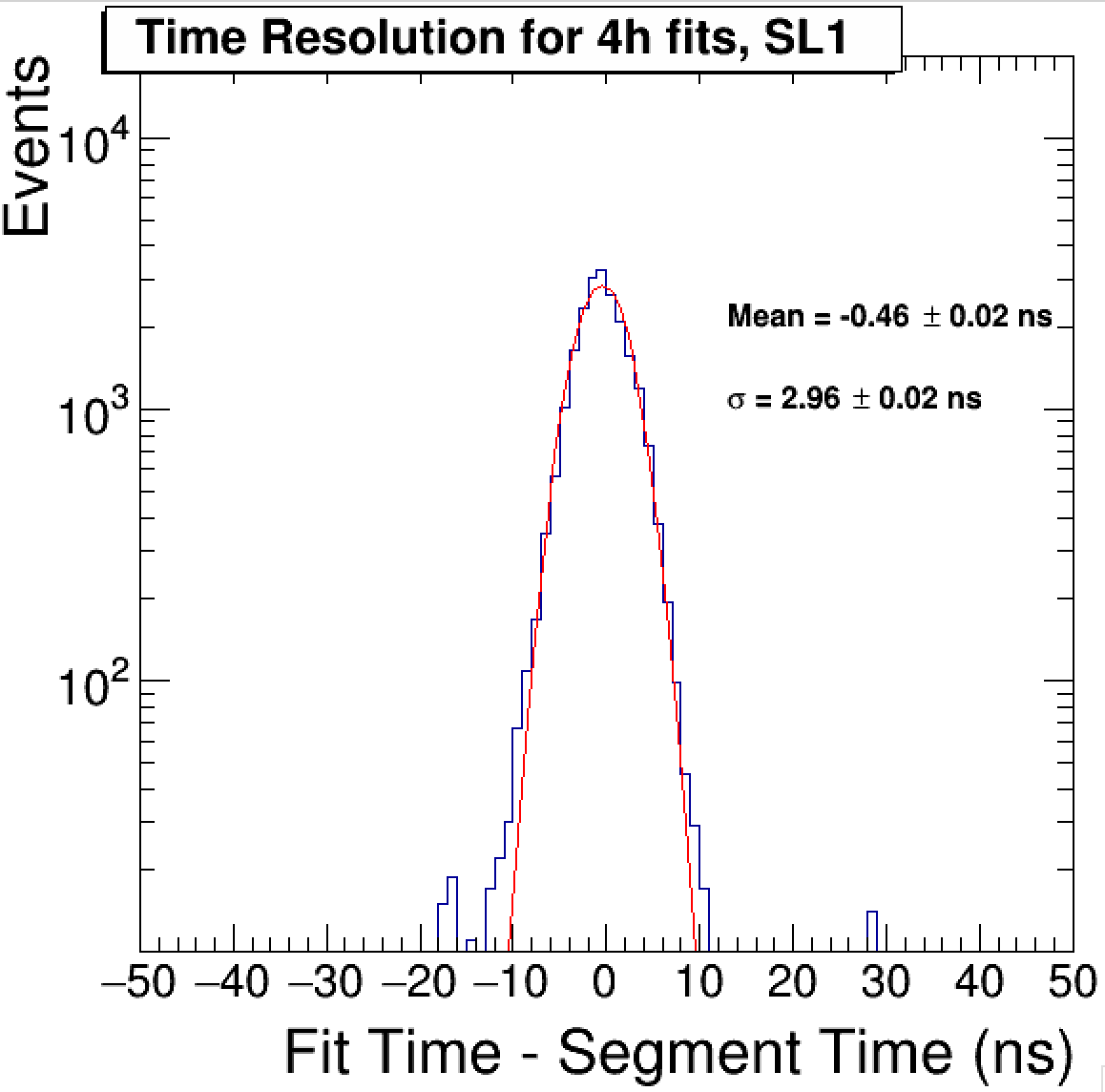}}
  \qquad
  \subfloat[]{\includegraphics[width=0.4\linewidth]{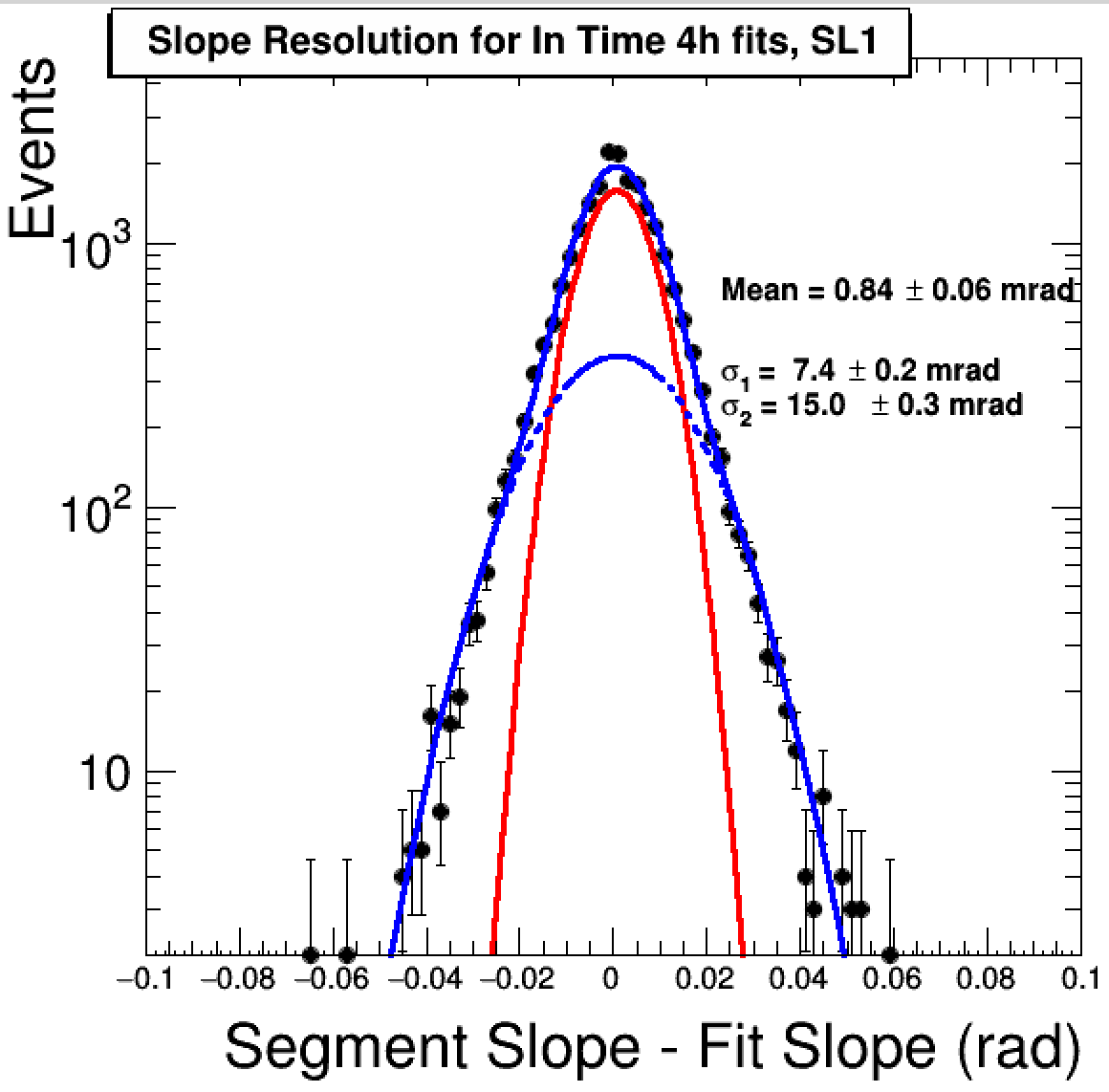}}
  \caption{ Time (a) and slope (b) resolutions for 4 hits fits in SL1}
  \label{fig:fig_timeslope_4h_int}
\end{figure}

Figure~\ref{fig:fig_timeslope_4h_int}(a) shows the time difference between the trigger primitive and the offline segment (referred to as 'time resolution') fitted to a Gaussian for trigger primitives with 4 hits in superlayer 1. The sigma of the Gaussian fit is $\sim$ 3~ns. Figure~\ref{fig:fig_timeslope_4h_int}(b) shows the difference between the trigger primitive reconstructed slope and the offline segment reconstructed slope (referred to as 'slope resolution'), fitted to two Gaussians, for primitives with 4 hits in superlayer 1. The sigma of the narrow Gaussian fit is $\sim$ 7~mrad. Because the offline segment is reconstructed from hits in more than one superlayer, the corresponding time and slope measurements constitute a good proxy for 'true' value of the considered variables, and only a small correlation is expected with the trigger primitive measurements, which at this level consider information from a single superlayer only.

\begin{figure}[!htb]
  \centering
  \subfloat[]{\includegraphics[width=0.4\linewidth]{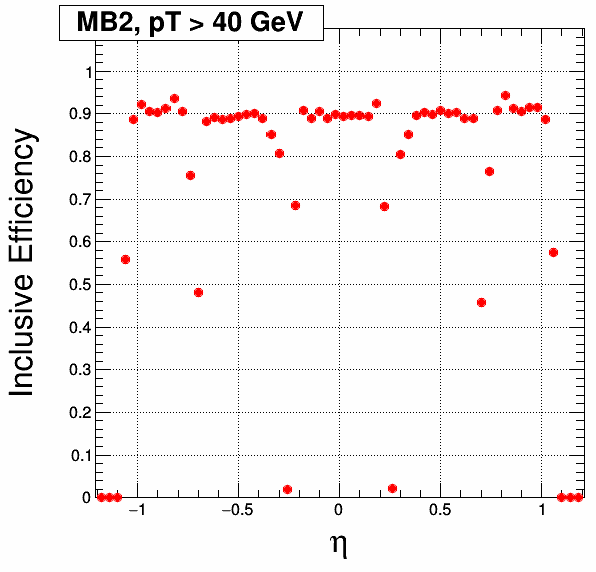}}
  \qquad
  \subfloat[]{\includegraphics[width=0.4\linewidth]{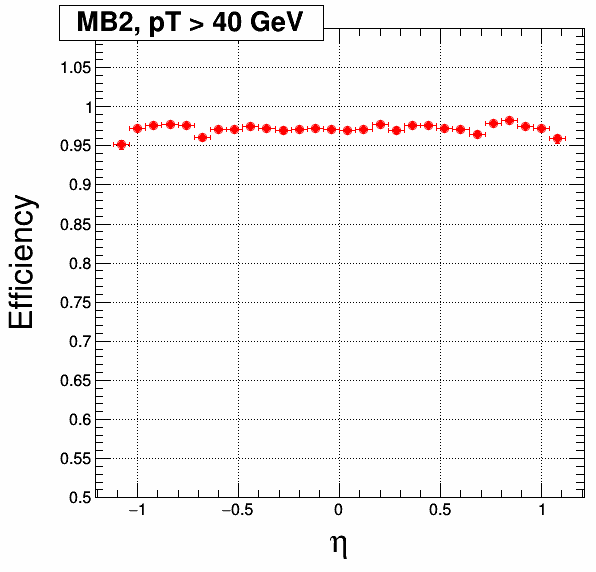}}
  \caption{Incluisve intrinsic efficiency (a) and efficiency with respect to offline segments with 4 hits (b)}
  \label{fig:fig_eff_int}
\end{figure}

Figure~\ref{fig:fig_eff_int}(a) shows the inclusive trigger primitive efficiency in Station 2, computed on a Monte Carlo muon sample with zero pileup (i.e., no other simultaneous interactions in the collision bunch crossing), for muons with pt $>$ 40 GeV. The trigger primitive is reconstructed from hits associated to the offline segment. All track parameters, including time, are obtained from a least square fitting method. The denominator is made of all muons passing a set of standard quality cuts. For the numerator we require at least one fit of any quality (i.e., fits with either 3 hits or 4 hits) in SL1 or in SL3 in the right BX. The correlated fit is used when available. The measured efficiencies are high, reflecting very good time resolution in all chambers. Observed dips are related to geometrical acceptance. When requiring an offline segment with more than 4 hits in that chamber, efficiencies are $>95\%$ everywhere, as shown in Figure~\ref{fig:fig_eff_int}(b)

\section{Performance of  the Analytical Method as implemented in the Emulator}
\label{EmuPer}
\begin{figure}[!htb]
  \centering
  \includegraphics[width=0.4\linewidth]{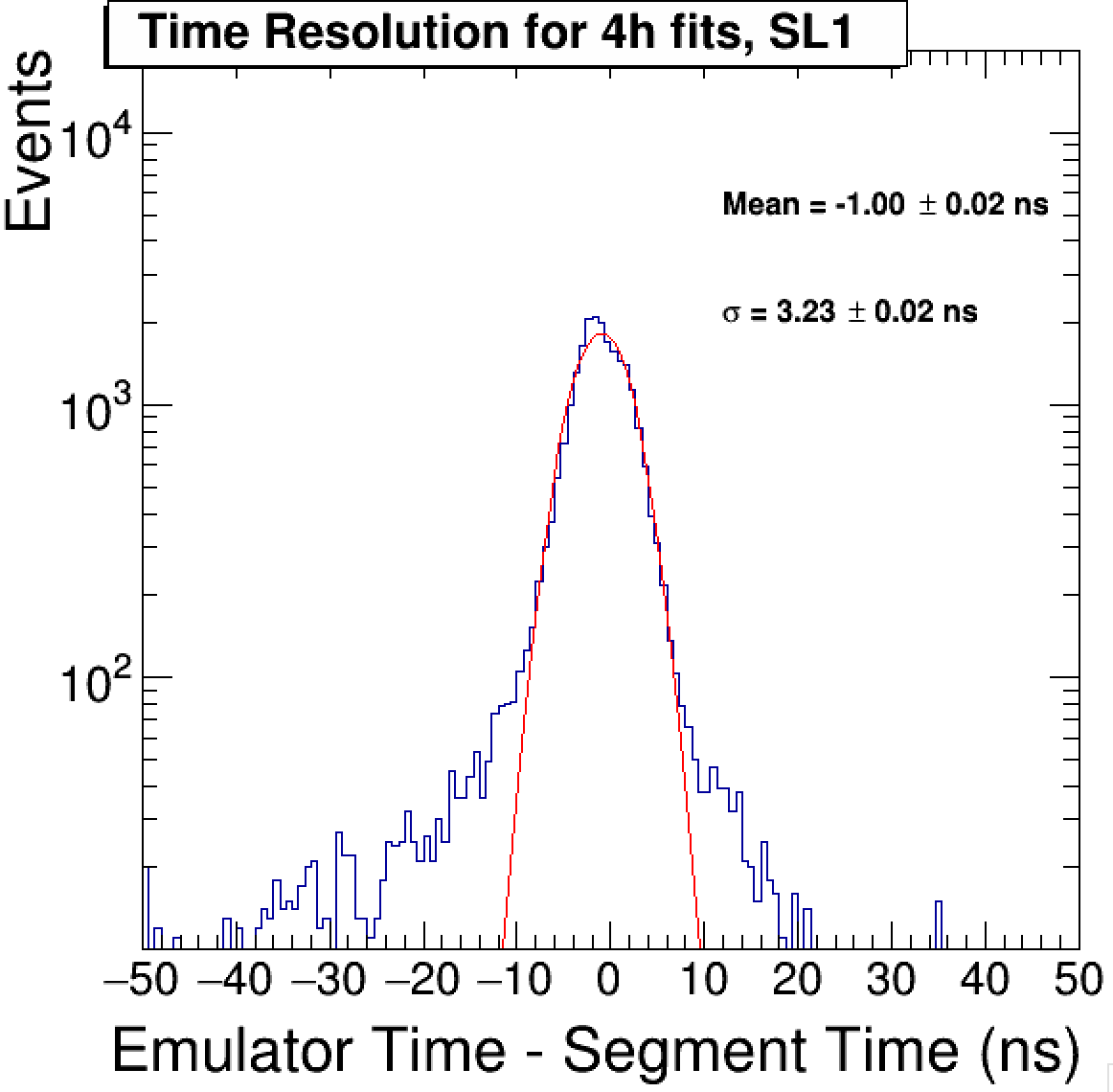}
  \caption{Time resolution for fits of 4 hits in SL1, as obtained by the Emulator.}
  \label{fig:fig_time_4h_emu}
\end{figure}

Figure~\ref{fig:fig_time_4h_emu} shows the difference between trigger primitive reconstructed time and offline segment reconstructed time ('time resolution'), fitted to a Gaussian, for primitives with 4 hits in superlayer 1. The trigger primitive is reconstructed with the Analytical Method as implemented in the software emulator (including full grouping and pattern recognition, and where the time for fits of 4 hits is calculated as arithmetic mean of 4 combinations of 3 layers), for muons coming from Z bosons decays in a 2016 sample. The sigma of the Gaussian fit is $\sim$3~ns. Tails appear in the distribution, in comparison with results when the input is given only by hits associated to offline segment in section~\ref{IntPer}

\begin{figure}[!htb]
  \centering
  \subfloat[]{\includegraphics[width=0.4\linewidth]{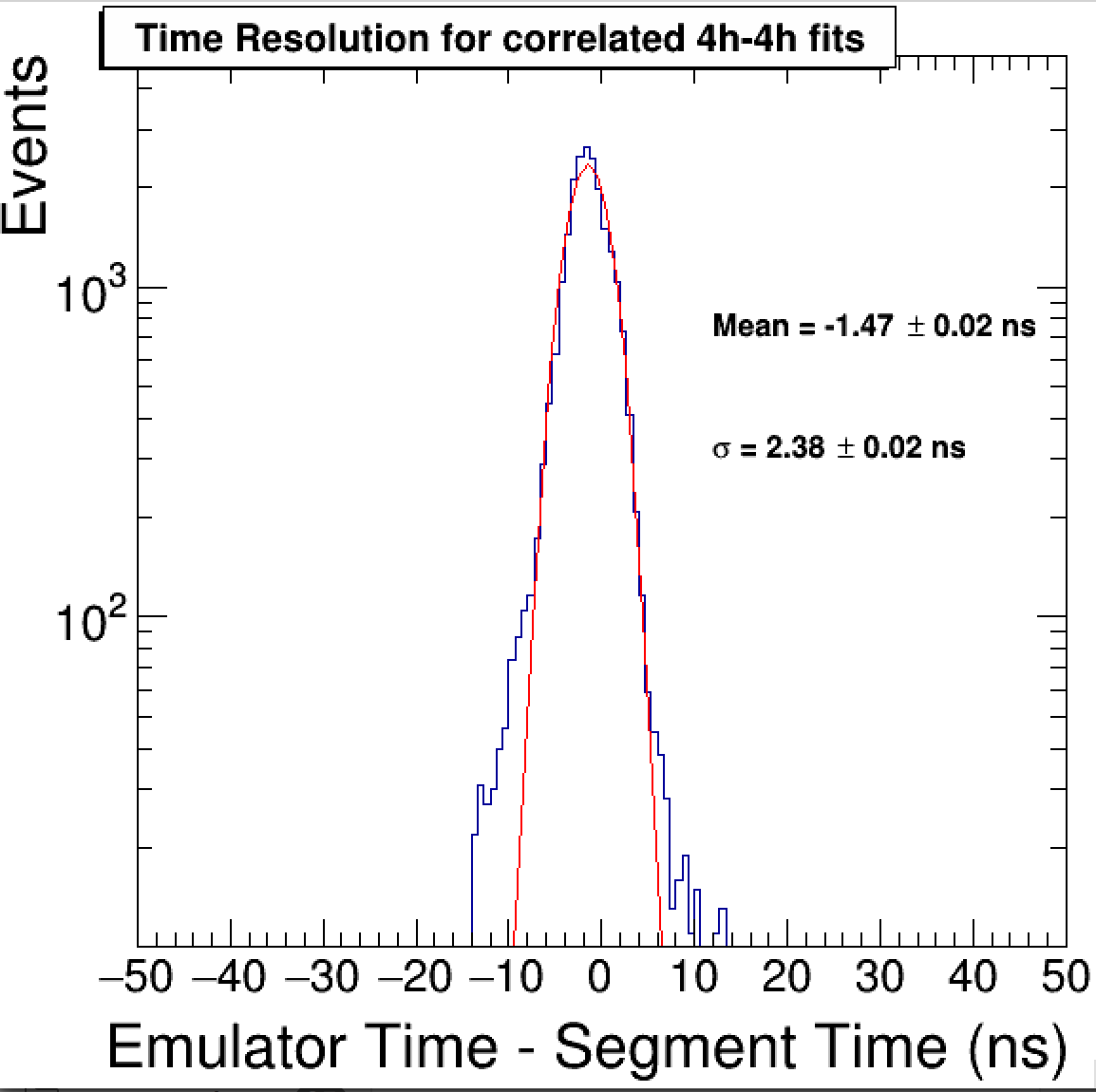}}
  \qquad
  \subfloat[]{\includegraphics[width=0.4\linewidth]{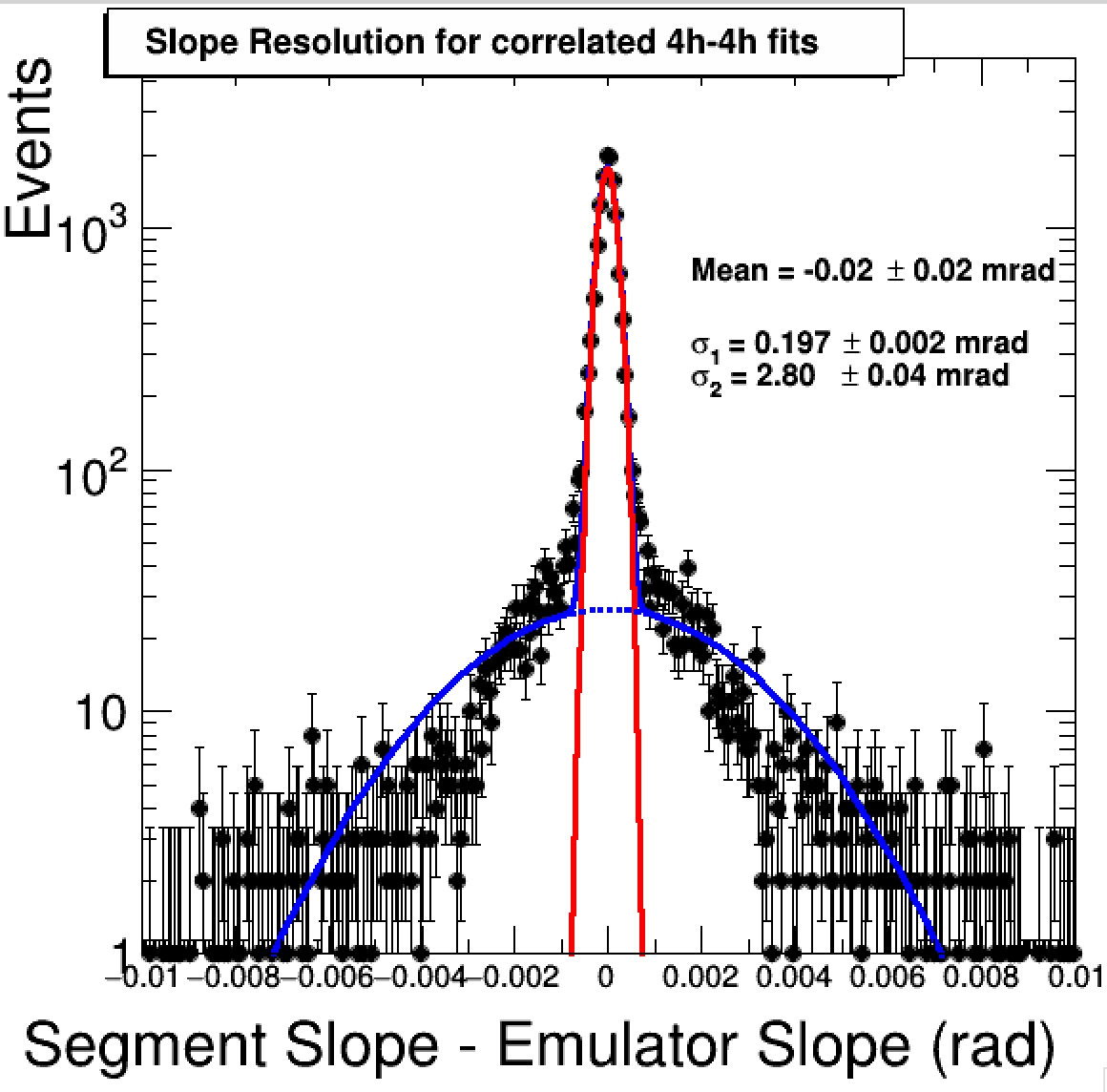}}
  \caption{Time (a) and slope (b) resolutions for correlated 4 hits - 4 hits fits, as obtained by the Emulator.}
  \label{fig:fig_timeslope_cor_emu}
\end{figure}

Figure~\ref{fig:fig_timeslope_cor_emu}(a) shows the difference between trigger primitive reconstructed time and offline segment reconstructed time ('time resolution'), fitted to a Gaussian, for correlated primitives with 4 hits in superlayer 1 and 4 hits in superlayer 3. The sigma of the Gaussian fit is $<$3~ns. A reduction of tails with respect to results for single superlayer is noticeable, due to the use of information from two superlayers.
Figure~\ref{fig:fig_timeslope_cor_emu}(b) shows the difference between trigger primitive reconstructed slope and offline segment reconstructed slope ('slope resolution'), fitted to two Gaussians, for correlated primitives with 4 hits in superlayer 1 and 4 hits in superlayer 3. The sigma of the narrow Gaussian fit is $<$1~mrad. The large improvement with respect to fits in a single superlayer comes from the increased lever arm given by the $>$20~cm distance between SL1 and SL3. Relevant correlations between emulator results and variables obtained from offline segment are possible in this case, because similar information is used for the calculations.

\section{First test data results}
\label{DQM}
\begin{figure}[!htb]
  \centering
  \includegraphics[width=0.6\linewidth]{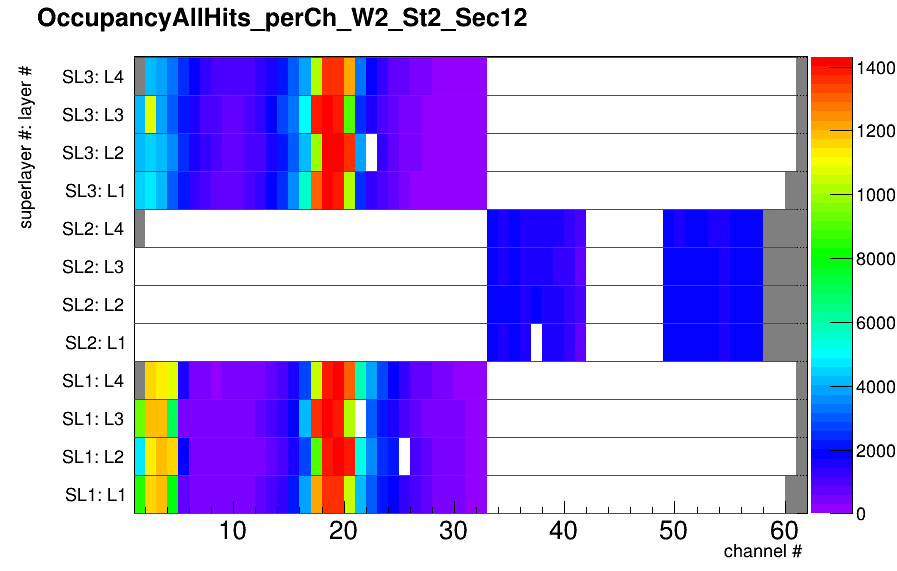}
  \caption{Occupancy of muon hits recorded in the legacy chain, and triggered by the phase 2 electronics chain, as shown by the DT Data Quality Monitor application. The distribution shows that the new trigger is working as expected: hits are read out by the legacy chain when triggering on the upgrade chain. }
  \label{fig:DQMOccup}
\end{figure}

In a spare DT chamber signals have been split into the two different electronics chains such that the legacy electronics chain and the upgrade chain register the same cosmic ray muon event. Figure~\ref{fig:DQMOccup} shows the occupancy of muon hits recorded in the legacy chain, and triggered by the phase 2 electronics chain implementing the Analytical Method algorithm for trigger primitive generation, as shown by the CMS DT Data Quality Monitor application. In each superlayer half of the cells signals are split to the phase 2 chain and read out by the legacy electronics: cells 1-30 in the phi superlayer and cells 31-57 in the theta superlayer. The frontend signals of cells 1-4 of SL1 and of cells 17-20 of SL1 and SL3 are used by the phase 2 electronics to produce trigger primitives. Figure~\ref{fig:DQMOccup} shows the occupancy of hits readout by the legacy electronics: the distribution shows that the new trigger is working as expected. For triggers from channels 1-4 in superlayer 1, hits are found on superlayer 3, but with low efficiency due to rough timing. Superlayer 2 was not fully commissioned at the time of the test.


 

\section{Conclusions}

 A proposal for an algorithm to generate DT Trigger primitives for phase 2 has been developed and implemented in firmware. Initial studies on CMS data and simulation show that a performance comparable to current offline reconstruction can be achieved at L1 trigger in Phase 2. The firmware implementation was tested using: (i) data from cosmic ray muons, and (ii) injection of muon data from Z decays. Good results have been obtained so far.





\end{document}